**Tracking picosecond strain pulses in heterostructures that exhibit giant magnetostriction**

S. P. Zeuschner[1,5], T. Parpiiev[2], T. Pezeril[2], A. Hillion[3], K. Dumesnil[3], M. Anane[4], J. Pudell[5], L. Willig[5], M. Rössle[1], M. Herzog[5], A. von Reppert[5]*, M. Bargheer[1,5]

[1]*Helmholtz-Zentrum Berlin, Wilhelm-Conrad-Röntgen-Campus, BESSY II, Albert-Einstein-Straße 15, 12489 Berlin, Germany*
[2]*Institut des Molécules et Matériaux du Mans (UMR CNRS 6283), Université du Maine, 72085 Le Mans cedex, France*
[3]*Institut Jean Lamour (UMR CNRS 7198), Université de Lorraine, 54000 Nancy, France*
[4]*Unité Mixte de Physique CNRS/Thales (UMR CNRS 137), 91767 Palaiseau, France*
[5]*Institute of Physics and Astronomy, University of Potsdam, Karl-Liebknecht-Straße 24-25, 14476 Potsdam, Germany*
\* reppert@uni-potsdam.de

We combine ultrafast X-ray diffraction (UXRD) and time-resolved Magneto-Optical Kerr Effect (MOKE) measurements to monitor the strain pulses in laser-excited $TbFe_2$/Nb heterostructures. Spatial separation of the Nb detection layer from the laser excitation region allows for a background free characterization of the laser-generated strain pulses. We clearly observe symmetric bipolar strain pulses when the excited $TbFe_2$ surface terminates the sample and a decomposition of the strain wavepacket into an asymmetric bipolar and a unipolar pulse, when a $SiO_2$ glass capping layer covers the excited $TbFe_2$ layer. The inverse magnetostriction of the temporally separated unipolar strain pulses in this sample leads to a MOKE signal that linearly depends on the strain pulse amplitude measured through UXRD. Linear chain model simulations accurately predict the timing and shape of UXRD and MOKE signals that are caused by the strain reflections from multiple interfaces in the heterostructure.

**1) Introduction**

The generation, propagation and detection of laser induced strain waves contain rich physics that has been studied extensively since the seminal work of Thomsen et al. that exploited photoelasticity for detection.[1,2] Ingenious all-optical probing schemes for GHz to THz phonons have since then been used to investigate strain waves in multiple materials,[3–6] vibrational modes of nanoparticles,[7] shear waves,[8] nonlinear propagation effects[9,10] and acoustic solitons.[11,12] Strain waves that originate from coherent phonon excitation can attain transient stresses on the order of GPa, which have been shown to interact with other phenomena such as phase transitions,[13,14] quantum well bandgaps,[15] piezo-[16]/ferroelectricity[17] and magnetism.[18–21] Such interactions are not only of fundamental interest but may also become relevant for applications as soon as the understanding allows for controllability.[22,23]

In order to study the response to pure strain pulses it is beneficial to spatially separate the laser excited transducer from the probed layer. The inherent limitation given by the finite optical penetration of the visible light in the transducer is often circumvented by backside probing schemes.[10,21] The development of (sub)-picosecond hard X-ray diffraction has opened the possibility to directly obtain the time-resolved strain amplitude[24–27] with penetration depths in the few µm regime. The separation of the layer peaks in reciprocal space allows for material specific probing of the energy flow[28,29] and strain evolution[30,31] in nanoscopic, crystalline heterostructures.

The envisioned manipulation of the polarization and magnetization states in ferroic materials via strain relies on a strong coupling between spin or electronic degrees of freedom and the atomic lattice.[22,32] In this regard, rare earth based alloys such as Terfenol (TbFe$_2$) have attracted attention due to the discovery of "giant magnetostriction"[33,34] i.e. lattice strains in excess of $10^{-3}$ caused by magnetization change. Among the binary rare-earth alloys TbFe$_2$ exhibits the largest magnetostriction.[35] It combines the large exchange interaction of the 3d orbital in Iron (Fe) with the large magnetic moment of 9$\mu_B$ per Tb atom, and the large spin-orbit coupling associated to 4f orbitals. The exchange coupling results in a ferrimagnetic alignment of the Fe and Tb moments with the Curie point ($T_C \approx$ 700 K) considerably above room temperature.[36] TbFe$_2$ crystallizes in a cubic C$_{15}$ laves phase structure, where the <111>-direction is the magnetic easy axis, which can be modeled by the cubic crystalline anisotropy constants $K_1$= -1.2·10$^8$ erg/cm$^3$ and $K_2$=2.08·10$^7$ erg/cm$^3$.[37] The resulting high coercivity is often reduced for application purposes by introducing Dy (<001> easy axis) to obtain the ternary alloy Terfenol-D (Tb$_x$Dy$_{1-x}$Fe$_2$). The desired low coercivity with large magnetostriction that is favorable for magneto-acoustic transduction applications can be tailored by different ratios x, where x = 0.27 is found to be optimal at room temperature.[38] Despite the potentially rich, coupled magnetization and lattice dynamics in this magnetostrictive ferrimagnet, there have been only few reports[8,22] that aim at quantifying and correlating the strain evolution and its coupling to the magnetization by time-resolved measurements in binary rare earth alloys.

Here we display the different capabilities of table-top ultrafast X-ray diffraction (UXRD) and all-optical methods to probe the strain propagation and evolution in a layered magnetostrictive heterostructure. Femtosecond laser pulses are used to excite the rare earth alloy Terfenol (TbFe$_2$), which serves as a transducer for strain waves into adjacent layers. UXRD measurements observe the arrival and shape of the strain waves in a thin, buried detection layer. From this, we extract the stress profile that generates the strain wave in the inhomogeneously excited TbFe$_2$ layer. The timings of the observed experimental features are rationalized by modeling the strain propagation in this multilayer sample using a 1D linear chain model of masses and springs. The modeling is shown to be particularly useful when the transducer is capped by a transparent layer so that the conventional symmetric bipolar strain pulse is split into an asymmetric bipolar pulse travelling into the transducer and a unipolar strain pulse that is reflected at the sample-surface after a time determined by the transparent layer thickness. Complementary to the UXRD data we employ an all-optical polarization sensitive measurement that probes the strain propagation in the transparent silica (SiO$_2$) acoustic delay line. Time-resolved magneto-optical-Kerr-effect (MOKE) measurements are shown to be a very sensitive probe for the arrival of the multiple strain echoes at the top of the laser-excited TbFe$_2$ layer, that can be used to complement the bulk sensitive UXRD.

**2) Experimental details**

We investigate laser excited samples that consist of (110) oriented Terfenol (TbFe$_2$) layers grown by MBE on (11-20) oriented Sapphire (Al$_2$O$_3$) with a buried Niobium (Nb) (110) buffer layer as previously described.[37,39] The basic sample structure is only capped by a 2 nm thin protective Titanium (Ti) layer, which does not significantly contribute to the experimental transients. We therefore refer to sample 1 as uncapped. The second sample was instead capped with an 885 nm thick amorphous silica (SiO$_2$) layer. The UXRD measurements are carried out at a laser-driven, plasma-based diffraction setup (PXS) that supplies 200 fs X-ray pulses at Cu $K_\alpha$- energy.[40] The table-top laser-pump x-ray-probe setup uses p-polarized excitation pulses at a central wavelength of 800 nm, with 1 kHz repetition rate and a full width at half maximum spot size of a 2-dimensional Gaussian function of 1.4 x 1.5 mm for the laser pulses and 0.3 x 0.3 mm spot size of the X-ray pulses. Using the top-hat approximation with the 1/e width for the laser excitation profile and the pulse energy, we calculate the fluence for the $TbFe_2$ and Nb experiments, respectively. The pump-fluence at the Nb angle is approximately 4% larger compared to the $TbFe_2$ experiments since the larger diffraction angle leads to a smaller laser footprint.

A representative reciprocal-space map (RSM) of the uncapped sample structure obtained at the PXS alongside the static X-ray diffraction curve and the temporal evolution of the material specific Bragg peaks are displayed in Fig. 1. In the probed RSM volume we find three separated peaks with their maximum intensity at the out-of-plane reciprocal space coordinate $q_z$ = 2.42 Å$^{-1}$, 2.64 Å$^{-1}$ and 2.69 Å$^{-1}$, that are attributed to $TbFe_2$ (220), $Al_2O_3$ (11-20) and Nb (110) respectively, according to their bulk lattice plane spacings. The layer thicknesses set by the sample growth are 500 nm $TbFe_2$ on top of 50 nm Nb and 330 μm $Al_2O_3$ as schematically depicted in Fig. 1e).

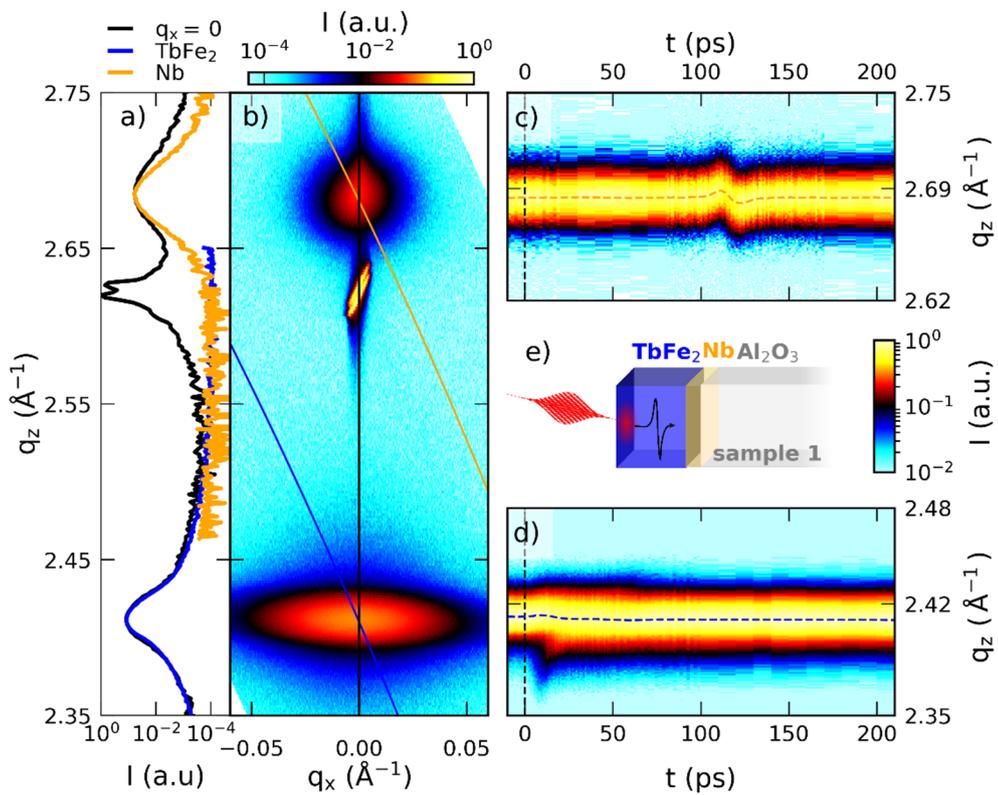

**Figure 1: Characterization of sample 1 via X-ray diffraction:** a) Slice of the reciprocal space map shown in b) at $q_x$ = 0 (black line). The blue and orange line in a) correspond to the probed reciprocal slice when using the convergent beam of the X-ray focusing optic and area detector at the lab-based diffraction setup at a fixed angle of incidence. c) and d) depict the temporal evolution of the Nb and $TbFe_2$ peak at 13.3 mJ/cm$^2$, respectively, with the fitted peak position indicated by dashed lines. e) Schematic depiction of the uncapped sample structure.

The $TbFe_2$ diffraction peak is significantly broadened along the in-plane reciprocal space coordinate $q_x$ compared to the instrument function limited $Al_2O_3$ substrate peak. This is a hallmark for microcrystalline domains that in this case exhibit a large mosaic spread of 1.5° around the bulk diffraction angle,[39] very similar to previously reported UXRD experiments on ferroelectric samples.[41,42] The presence of such structural imperfections in the $TbFe_2$ becomes evident by comparison to the diffraction peak of the Nb layer which exhibits a much smaller width in $q_x$. The blue and orange lines in Fig. 1b) represent the reciprocal space slices that are probed in our setup for two fixed angles of incidence (AOI) ω that are chosen to be selectively sensitive to the $TbFe_2$ and Nb lattice strains respectively.

The combination of an optic that focuses X-rays onto the sample with a convergence of Δω≈0.3° (Montel optic from Layertec) and an X-ray area pixel detector (Dectris PILATUS-100k) allows for swift data acquisition that avoids time-consuming mesh scans of the AOI (ω) and the diffraction angle (θ). Each pixel of the X-ray area detector is mapped to reciprocal space coordinates $q_x$ and $q_z$ using the mapping routine described in a previous publication,[41] which is applicable in the thin film regime. Using this fixed angle detection scheme the flux of $10^6$ photons/sec incident on the sample is sufficient to probe the evolution of the material specific diffraction peaks, of the laser excited heterostructure with subpicosecond time resolution within few hours. The X-ray diffraction curves of the unexcited sample at the Nb and TbFe$_2$ AOI are indicated as orange and blue solid lines in Fig. 1a) and their time evolution is represented by Fig. 1c) and 1d) for Nb and TbFe$_2$, respectively. The dashed lines in Fig. 1c) and 1d) indicate the temporal evolution of the peak center that is extracted by fitting the diffraction signal with a Gaussian line profile at each delay. This extracted Bragg peak position in reciprocal space is inversely proportional to the lattice constant $d$ of the material via $q_{z,\text{Fit}}(t) = \frac{2\pi}{d_{\text{Fit}}(t)}$. UXRD thus probes the time-resolved strain $\varepsilon(t)$, defined as the change of the average lattice constant $d$ relative to the unexcited sample $\varepsilon := \frac{d(t)-d(t<0)}{d(t<0)}$.

### 3) Analysis and discussion of the results

### 3a) Signatures from the sample without SiO$_2$ capping

Before discussing the transient strain of the SiO$_2$ capped Terfenol (TbFe$_2$) structures, it is instructive to rationalize the signals seen in the UXRD experiment on the uncapped sample 1 that is schematically depicted in Fig. 1e) for a fluence of 12.7 mJ/cm² and 13.3 mJ/cm² for the TbFe$_2$ and Nb, respectively. At first we discuss the strain evolution in the directly excited, approximately 500 nm thick TbFe$_2$ layer shown in Fig. 2a). The blue data points show the experimentally obtained strain from Gaussian fits to the diffraction curves. Representative fits and raw time-resolved data from Fig. 1c) and d) are shown in Fig. S1 of the supplementary information. Beyond 20 ps we observe an expansion, that becomes manifest in a shift of the diffraction peaks to smaller $q_z$. Within the first 20 ps one observes a transient shift of the majority of the Bragg peak to larger $q_z$, which coincides with the appearance of a shoulder at smaller $q_z$. Between 40 and 80 ps we detect a pronounced, triangular shaped strain increase and subsequent decrease in addition to an overall rising background.

No background is observed in the strain response of the 50 nm thin, buried Nb layer displayed in Fig. 2b). The strain in the Nb layer is close to zero up to 4 nanoseconds (not shown), except for the very pronounced, nearly symmetric bipolar strain pulse that starts with a contraction at approximately 90 ps, reverses sign at 117 ps and ceases at approximately 150 ps. The diffraction peak evolution of the thin Nb layer displayed in Fig. 1c) exhibits a large peak shift that corresponds to a strain amplitude of 1.5 ‰, which is only present during a short time window. Inset c) in Fig. 2. shows the strain oscillation in Nb for different pump fluences, normalized to the fluence of the laser pulses. Since the transient strain curves nearly coincide, our data evidece a linear fluence dependence for this strain oscillation feature in Nb up to 13.3 mJ/cm².

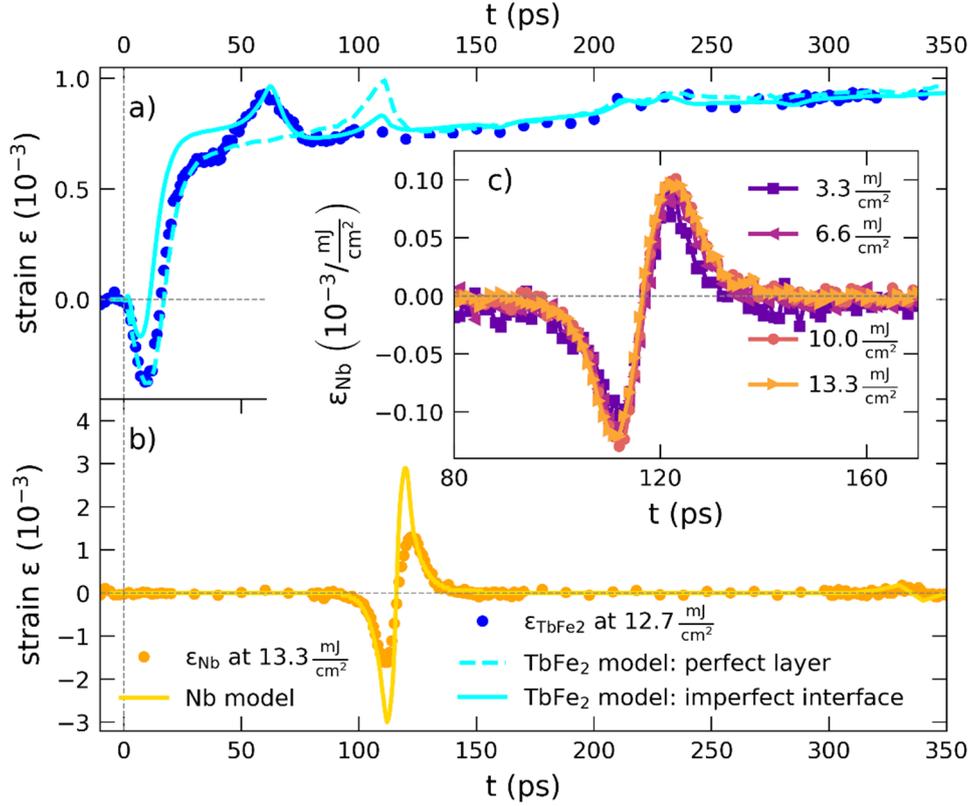

**Figure 2: Transient strain signatures of sample 1 without SiO₂ capping:** a) and b) display transient strains (symbols) extracted from the average peak shift via Gaussian lineshape fits (see Fig. S1 of the supplementary information) and the simulated strain response (solid lines) using the udkm1Dsim toolbox. The dashed line in a) corresponds to a model with a full single-crystalline TbFe₂ layer, whereas the solid line takes a disordered TbFe₂ layer at the TbFe₂/Nb interface into account. Inset c) depicts the normalized transient strain in the Nb layer for different excitation fluences.

The interpretation of the UXRD data from the thick TbFe₂ layer is based on the insights on ultrafast lattice response of photoexcited thin films studied by UXRD that were previously discussed by Schick et al.[43] The laser illumination leads to the excitation of coherent and incoherent phonons, which superimpose in the strain response of the absorbing layer.[2] The strain pulse composed by the coherent excitation of phonons subsequently propagates at the longitudinal acoustic phonon velocity, whereas the thermal energy leaves the excited layer by a slower diffusion process.

Our experiment represents the limiting case of an inhomogeneously excited transducer since the TbFe₂ layer thickness is approximately 25 times larger than the 19.9 nm light intensity penetration depth at 800 nm, that we obtain from ellipsometry measurements using a commercial setup and analyzing software (SENTEC), as discussed in section 3 of the supplementary. The appearance of a marked shoulder in the TbFe₂ diffraction signal on the lower $q_z$ side for the main diffraction in Fig. 1d) signals the existence of a highly strained surface layer on top of the nearly unperturbed TbFe₂. Schick et al.[43] have analyzed in detail that the exponential stress profile originating from inhomogeneous laser heating leads to an initial compression of the majority of the layer. Since the Gaussian fit is most sensitive to the central region of the diffraction peak we observe the leading compressive strain front as a shift of the Bragg peak maximum to larger angles in the first 15 ps. When the free surface expansion propagates into the material at the speed of sound the strong expansive component finally shifts the Bragg peak maximum to smaller angles. We attribute the remaining slope to the heat transport that equilibrates the inhomogeneous temperature profile within the TbFe₂ layer on a timescale of hundreds of picoseconds to several nanoseconds.

The strain response of the Nb layer seen in Fig. 2b) confirms that the thermal transport occurs mainly within the TbFe$_2$ layer since we observe no thermal expansion that would appear as a background within our 4 ns measurement window. The bipolar strain pulse marks the delayed passage of the coherently excited phonon wave packet, which is launched at the sample air interface, through the buried Nb layer. The detected diffraction peak shift of the 50 nm thick detection layer thus shows a background-free signal of the strain pulse, consisting of a compressive leading edge, which is followed by an expansive trailing edge as it is known from previous picosecond acoustic investigations.[2,44] The smaller layer thickness leads to higher average strain signals and sharper features as compared to the strain detected in the thick transducer layer.

The signature of the exit of the bipolar strain pulse from the probed TbFe$_2$ layer is an increase of the average layer strain followed by a decrease back to the thermal expansion background since the leading compressive edge exits while the trailing expansive part is still in the layer. In our experiment we observe a pronounced delay between the exit of the strain wave from the TbFe$_2$ layer, at approximately 40 ps and its arrival in the adjacent Nb at 90 ps. This 50 ps delay of the signatures can only be rationalized if the strain pulse traverses a TbFe$_2$ layer that does not contribute significantly to the X-ray diffraction signal. Using $v_{sound}$ = 3.94 nm/ps, known for polycrystalline TbFe$_2$[45], this corresponds to a layer with a thickness of approximately 187 nm TbFe$_2$ which has a considerably different texture. Structural inhomogeneities are in line with the mosaic peak broadening and the comparably small X-ray diffraction intensity of the TbFe$_2$ peak. The existence of a structurally imperfect interface layer at the TbFe$_2$-Nb interface is further supported by in-situ Atomic Force microscopy (AFM) measurements that observed that the rare-earth alloy layer growth proceeds first as separated 3-dimensional islands that only coalesce to form a continuous film for thicknesses on the order of 100 nm and above.[46] The large in-plane lattice-constant mismatch of 11.6% between the underlying Nb template and the TbFe$_2$ is reduced by a thin FeNb layer, but is probably the origin for the large mosaicity and for the limited coherence length along the growth direction[46] which amounts to 50 nm in the present TbFe$_2$ film.

Modeling the excitation and propagation of picosecond acoustic strain pulses is achieved by solving the partial differential equation for the time-dependent local strain in which the given spatio-temporal stress profile acts as source term.[2,43,47] The thin film geometry reduces this to a 1-dimensional problem as the laser excitation spot is much larger than the film thickness of the nanostructure and the probed X-ray spot. Numerical solutions for the strain evolution in nanoscopic layers are frequently applied in nanoscopic heterostructure geometries, where multiple interface reflections complicate analytical solutions.[31,47,48] In Fig. 2a) and b) we compare the UXRD data to simulation results obtained with the udkm1Dsim toolbox package that we used to calculate the time-resolved strain response based on a linear chain model (LCM) of masses and springs. Although details of the software are given in the reference publication,[49] we briefly outline the workflow of the used modeling. Upon input of the thermophysical material properties and the known sample geometry which are listed in Tab. 1 of the supplementary information, we first calculate the absorbed optical energy density and temperature profiles according to the heat diffusion equation[50] with unit cell resolution. The resulting spatio-temporal temperature profile represents the thermoelastic stress that drives a linear chain of masses and springs, where the masses represent individual unit cells. In the last simulation step the obtained time-resolved strains are used as an input for the computation of the time-dependent dynamical X-ray diffraction signal.[51] The resulting diffraction peaks are fitted with a Gaussian line profile to yield the strain signal displayed as lines in 2a) and b).

The dashed line in Fig. 2a) shows the modeled strain of a structurally perfect 436 nm thick TbFe$_2$ layer on top of a 50 nm Nb attached to an Al$_2$O$_3$ substrate. The simulation data represented by the solid line assumes only 249 nm structurally perfect TbFe$_2$ on top of a 187 nm TbFe$_2$ layer with substantial disorder. The improved fit of the model regarding the triangular feature beginning at 40 ps substantiates the evidence for a structurally different TbFe$_2$ layer at the Nb interface. The total TbFe$_2$ layer thickness is determined by the arrival time of the bipolar strain pulse in Nb using the directionally averaged speed of sound of $v_{sound}$ = 3.94 nm/ps for polycrystalline TbFe$_2$,[45] due to the lack of exact elastic constants for single crystalline TbFe$_2$. Despite the agreement between the simulated and experimental strain in the TbFe$_2$ layer, the simulation substantially overestimates the bipolar strain pulse amplitude in the Nb layer. This may be accounted for by taking into account a slowly rising stress profile in TbFe$_2$ and acoustic damping as well as via scattering of the coherent phonons in the interface layer.[52–54]

**3b) Signals in the SiO$_2$ capped sample**

The central experimental result of this work is summarized in Fig. 3. There we combine the experimental results from a sample capped by an amorphous SiO$_2$ layer as sketched in the inset 3e). Fig. 3a) shows the spatio-temporal strain profile that is obtained within the linear chain model for the second sample structure with the stacking sequence 882 nm SiO$_2$ / 342 nm TbFe$_2$/ 50 nm Nb/ Al$_2$O$_3$ substrate, which is solved by the udkm1Dsim toolbox[49]. Red and blue colors correspond to regions of expansive and compressive strain respectively. One observes that the expansion of the laser heated region slowly spreads as the heat diffuses within the TbFe$_2$ layer. Now, the bipolar strain pulse that is launched towards the substrate is clearly asymmetric, where a large amplitude leading compressive part is followed by a smaller expansive tail. In addition, a unipolar compression pulse propagates in the SiO$_2$ capping layer towards the surface, where it is converted into an expansion. The simulated strain in Fig. 3a) clarifies that for a very thin capping layer, the expansive unipolar wave reflected at the surface would superimpose with the small expansive tail of the asymmetric bipolar wave to form the symmetric bipolar wave observed in sample 1, see Fig. 2. Multiple reflections of the strain pulses occur at the material interfaces indicated by horizontal dashed lines, where the reflection and transmission occur according to the acoustic impedance.[2,55,56]

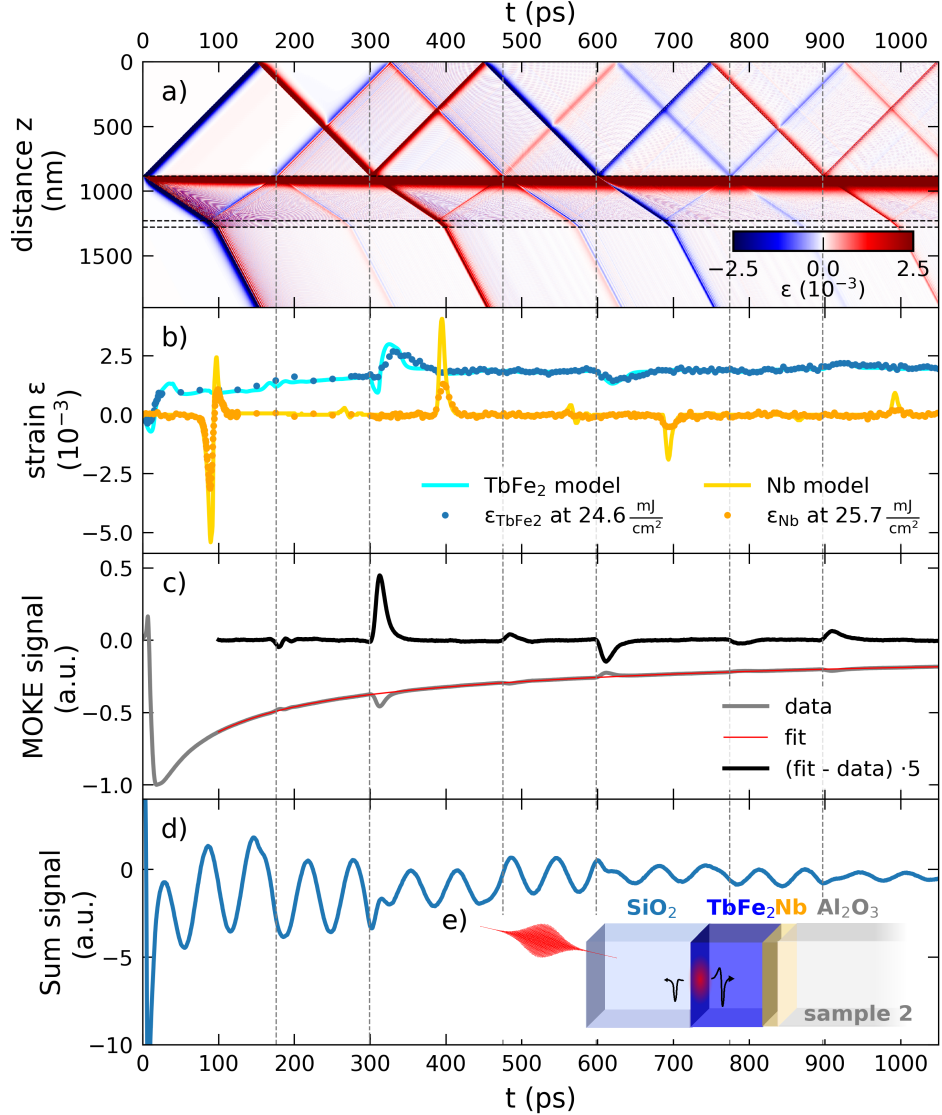

**Figure 3: Time-resolved signals from the SiO$_2$ capped sample structure:** a) Spatio-temporal strain simulation result that highlights the occurrence of multiple echoes from bipolar and unipolar strain pulses. b) Comparison of the strain signal from UXRD measurements and strain extracted from udkm1Dsim toolbox simulations. c) Time-resolved MOKE signal. Vertical dashed lines highlight the timings when strain pulses traverse the SiO$_2$/TbFe$_2$ interface according to the simulation. d) Time-domain Brillouin scattering signal from SiO$_2$. Inset e) schematically depicts the sample geometry in the experiment and the horizontal dashed lines in a) indicate the layer interfaces.

The occurrence of multiple unipolar strain pulse echoes that traverse the TbFe$_2$ and Nb layers at different timings is readily seen in the UXRD data presented in Fig. 3b). The modeled average strain shown as solid lines accurately predicts the timing and shape of the observed features but the amplitude of the Nb strain is substantially overestimated. This may indicate a finite electron-phonon-coupling time and scattering of the coherent phonons in TbFe$_2$ from structural imperfections, which are both not captured in the current modeling. The presented UXRD data were obtained under identical excitation conditions as the experiments on the uncapped sample except for the larger pump-fluence 24.6 mJ/cm², which leads to an increase of the detected strain amplitudes.

In the following we discuss the results of time-resolved MOKE measurements, which probe the change of the polarization state of the probe light upon reflection due to the permanent magnetization of the sample. The measurement displayed in Fig. 3c) was carried out close to the polar MOKE geometry with an external out-of-plane magnetic field of $\mu_0H$ = 800 mT using 200 fs laser pulses at a central wavelength of approximately 800 nm, a repetition rate of 250 kHz and a pump fluence of approximately 2.7 mJ/cm². In these measurements the difference of the polarization change for opposite external field orientations ($S(H_{up})$ - $S(H_{down})$) is probed using the reflection of 800 nm probe-light-pulses analyzed by a half-wave plate in combination with a Wollaston-prism and a balanced photo-diode. Lock-in detection using an acousto-optical modulation of the pump beam intensity at 50 kHz was employed. The resulting polar MOKE signal displayed as a solid grey line essentially probes the out-of-plane magnetization component of the $TbFe_2$ layer, within the 19.9 nm optical penetration depth and is probably sensitive to the Fe sub-lattice.[57] Subtraction of the slowly-varying thermal background approximated by a double-exponential decay (red line) from the MOKE signal (grey line) reveals multiple sharp peaks in the residual black curve. By comparison with the linear chain model results in Fig. 3a) it becomes obvious that the observed features occur at the time when the longitudinal strain pulse echoes traverse the top few nanometers of the $TbFe_2$ layer, given by the penetration depth of the probe pulse. Note that the sign of the peaks correlates with the sign of the (unipolar) strain-pulse echoes and that even the small reflections from the $TbFe_2$/Nb interface produce observable MOKE signatures at around 180, 485 and 790 ps.

Fig. 3d) displays the time-resolved polarization analysis signal of the reflected 800 nm probe beam independent of the magnetization state, which is obtained from the sum signal ($S(H_{up})$ + $S(H_{down})$) of the balanced detection. Similar time-resolved Brillouin scattering experiments have shown that the observed oscillations originate from the interference of the reflected light from the traveling strain pulse in the transparent $SiO_2$ medium and the static interfaces.[58,59] Pronounced phase jumps in this Brillouin signal occur when the strain pulses invert their sign due to the reflection at the $SiO_2$/air interface.[55,60]

Consequently, the strain propagation as modeled by the 1D-linear chain model accurately predicts the timings of all the experimental signals we observed in this heterostructure. We have employed a single temperature model for the driving stress on the lattice. This certainly oversimplifies the equilibration process of the electron-, lattice- and spin- subsystems to occur instantaneously. A detailed analysis of the dynamics prior to the equilibration is beyond the scope of the current investigation, as it requires the knowledge of the thermophysical properties for each sub-system as well as the coupling constants, of this largely unexplored material.

**3c) Experimental results from the buried detection layer**

In the henceforth presented data analysis we put the focus on the qualitative and quantitative information that can be directly extracted from the UXRD signal in the buried Nb detection layer. In Fig. 4a), we see that the normalized, background-subtracted MOKE signal originating from the $TbFe_2$, matches the normalized Nb strain, when shifted by 84 ps, which is the longitudinal acoustic propagation time through the $TbFe_2$ layer. This agreement proves a linear relation between the lattice strain and the observed MOKE signal. The slight discrepancies at 485 and 790 ps probably originate from the fact that the MOKE signal results from a superposition of the strain pulses reflected at the surface and at the $TbFe_2$/Nb interface, which traverse the top $TbFe_2$ layer simultaneously (see Fig. 3a). Since only part of the reflection at the $TbFe_2$/Nb interface is again reflected at the $TbFe_2$/$SiO_2$ interface, the Nb layer senses a different strain wave composition. It will be important for future experiments investigating the interaction of shear waves with the magnetization to accurately identify also the small longitudinal acoustic pulse echoes in such multilayered structures. Previous picosecond acoustics investigations in magnetic samples have observed that strain pulses can exert a torque on the sample magnetization $\vec{M}$ via a transient modification of the crystalline anisotropy, often resulting in a damped precessional motion of $\vec{M}$.[18,61,62] Although a torque on the magnetization by the strain pulse is expected, the absence of precessional oscillations challenges the theoretical interpretation of the observed MOKE signal based on the Landau-Lifshitz-Gilbert model. Crystalline defects in the $TbFe_2$, resulting in small magnetic domains,[63] in combination with a magneto-crystalline anisotropy and damping could drastically suppress the coherent precessional signal. The signal might have contributions from a modulation of the reflectivity driven via the photoelastic effect[21], however, the reflectivity signal does not exhibit significant spikes at the echo positions. In any case the striking resemblance of the detected strain pulses in the Nb layer to extracted features in the MOKE measurements demonstrates a high sensitivity of MOKE for probing strain pulses arriving at the $TbFe_2$ surface.

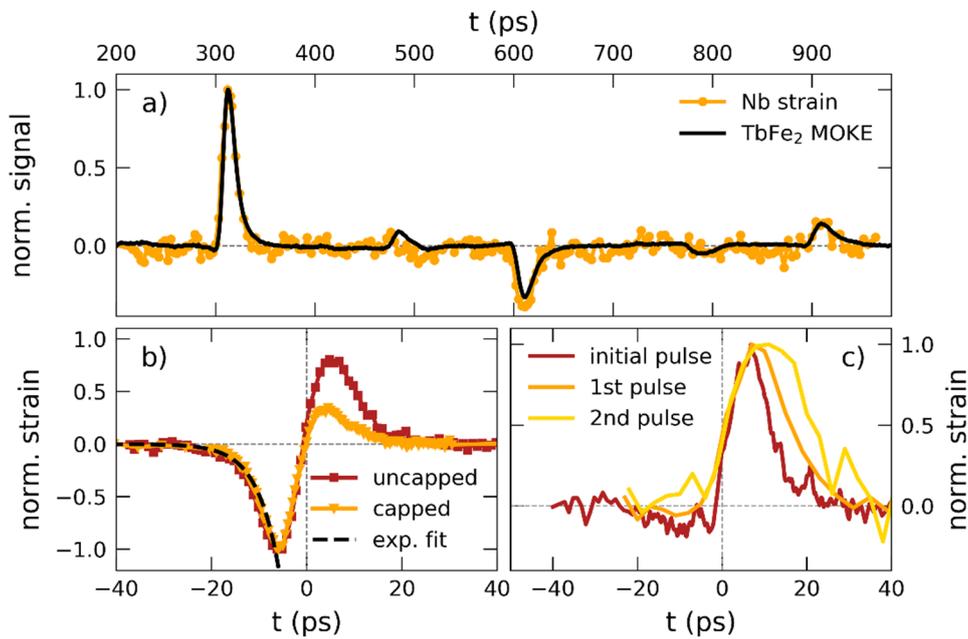

**Figure 4: Analysis of the strain pulse signatures:** a) Comparison of the coherent phonon strain contribution seen in the MOKE and UXRD signal, scaled to the maximum amplitude and shifted to overlap in time. b) Comparison of the initial asymmetric bipolar strain pulse in the capped sample 2 and the symmetric bipolar strain in the uncapped sample 1, to an exponential fit with a time-constant of 4.56 ps. c) Evolution of the strain pulse after passing the $SiO_2$ layer multiple times.

In Fig. 4b) we compare the initial bipolar strain pulses from the capped and uncapped samples, normalized to their compressive part. We observe that the leading, compressive parts coincide, whereas the expansive parts in the $SiO_2$ capped sample 2 is strongly reduced. The black dashed line indicates a single exponential fit to the falling edge of the compressive strain with a time constant of 4.6 ps, which translates to a spatial extension of approximately 18 nm using $v_{sound}$ = 3.94 nm/ps. This value provides an estimation of the spatial extend of the driving stress profile.[3,64] This matches the optical penetration depth obtained from ellipsometry, which shows that potential hot electron diffusion does not substantially increase the excitation profile length.

The evolution of the unipolar strain pulse within the $SiO_2$ capping is shown in Fig. 4c), where we compare the first and second unipolar strain echoes that traverse the Nb layer at 396 ps and 695 ps to the initial pulse launched into the $SiO_2$ capping. The latter is extracted from the difference between the bipolar strain pulses observed on the capped and uncapped samples. In this analysis we assume that the laser generated stress profile is identical in the capped and uncapped $TbFe_2$ samples, and that the difference of the initially detected bipolar strain pulse seen in Fig. 4b) originates solely from partial reflection of the expansion at the top $TbFe_2$ interface. The FWHM of the detected strain signals increases from 9 to 16 and 21 ps. The modeling does not include any broadening mechanisms and reports echoes with a constant width of 8 ps (FWHM). Anharmonic interactions in the lattice potential have been shown to change the shape and broaden high amplitude coherent phonon wavepackets.[65] Contributions from the $SiO_2$ surface-roughness should also be taken into account.

## 4) Conclusion

In this work we have combined multiple techniques to follow the trajectory of strain pulses that are generated by femtosecond laser pulses exploiting the giant magnetostriction material Terfenol ($TbFe_2$) as a transducer. MOKE measurements in $TbFe_2$ are shown to provide a surface sensitive method to probe strain pulses at the top of the metallic $TbFe_2$ that is complementary to bulk sensitive X-ray diffraction.

Probing the strain pulse in a thin, buried detection layer adjacent to an optically opaque transducer via UXRD allows for a characterization of the coherent strain pulse separately from heat expansion, without frontside-pump backside-probe schemes. By detecting the strain pulse in the buried and perfect Nb layer, we demonstrate how UXRD can clearly locate the structurally imperfect fraction of the $TbFe_2$ layer in this opaque heterostructure. Contrary to all-optical methods, UXRD provides a quantitative measure of the average lattice strain that does not require detailed knowledge of photo-elastic coefficients and optical properties. In combination with 1D-linear chain models, UXRD can provide quantitative information on realistic strains with unit cell resolution, that can be used as input for modeling strain-assisted magnetization switching approaches.[22] The obtained maximum strain amplitudes in $TbFe_2$ of $2\cdot10^{-3}$ is well below the deterministic switching limit of a few percent strain, but the local heating substantially lowers the anisotropy[66] as it is known from heat-assisted magnetic recording schemes.

The combination of MOKE and UXRD outlines a potential path towards an experimental calibration of the magneto-elastic coefficient i.e. the magnetization change per strain amplitude for picosecond acoustic pulses. Such a quantity is not only relevant for testing fundamental research that models magneto-elastic couplings but also represents a valuable input for application-oriented research.

We believe that probing the strain-pulse in a buried detection layer is a versatile method for studying the stress generation profile as it separates coherent from incoherent phonon excitations. It will be especially useful in situations where multiple mechanisms with different spatial or temporal characteristics superimpose in the strain generation process as it is the case in (anti-)ferromagnetic[64,67,68] materials. The use of an acoustic delay line further introduces the possibility to study the evolution of the strain pulse shape and to calibrate the magnetization response to unipolar compression and expansion pulses. This will support important future steps towards a full understanding of the demagnetization process especially in high-anisotropy, giant magnetostriction materials. A combination of time-resolved probes that monitor different degrees of freedom within the same experiment will foster the understanding of the intricate couplings between electron-, spin- and lattice systems in solids, which forms the basis for many useful devices.

## 5) Acknowledgement

We gratefully acknowledge the BMBF for the financial support via 05K16IPA and the DFG via BA 2281/8-1, BA 2281/11-1 and Agence Nationale de la Recherche under grant ANR-14-CE26-0008. We are thankful to Marwan Deb and Gregory Malinowski for stimulating discussions.